\documentclass[sigconf, nonacm]{acmart}
\newcommand{\eat}[1]{}
\usepackage{latexsym}
\usepackage{amsfonts}
\usepackage{amsmath}
\usepackage{color}
\usepackage{colortbl}
\usepackage{epsfig}
\usepackage{xspace}
\usepackage{graphicx}
\usepackage{subfigure}
\usepackage{pifont}
\usepackage{bm}

\usepackage{xparse}

\usepackage[lined,boxed,vlined,ruled,linesnumbered]{algorithm2e}
\usepackage{algorithmicx}
\usepackage{paralist}
\usepackage{enumerate}
\usepackage{ifthen}

\usepackage{makecell}
\usepackage{listings}
\usepackage{enumitem}

\newcommand{\stitle}[1]{\vspace{1.0ex}\noindent{\bf #1}}

\usepackage{epsfig}
\usepackage{multirow}
\usepackage{url}
\usepackage{float}
\usepackage[color,matrix,arrow,all]{xy}
\usepackage{soul}
\usepackage{tikz}
\usetikzlibrary{shapes,snakes}
\usetikzlibrary{calc}

\newtheorem{example}{Example}

\NewDocumentCommand{\ff}{ mO{} }{\textcolor{red}{\textsuperscript{\textit{ff}}\textsf{\textbf{\small[#1]}}}}

\NewDocumentCommand{\cc}{ mO{} }{\textcolor{violet}{\textsuperscript{\textit{cc}}\textsf{\textbf{\small[#1]}}}}
 
\newcommand{\sys}{\text{ReCoGNN}\xspace}

\newcommand{\bi}{\begin{itemize}}
	\newcommand{\ei}{\end{itemize}}

\newcommand{\be}{\begin{enumerate}}
	\newcommand{\ee}{\end{enumerate}}
\newcommand{\beqn}{\begin{eqnarray*}}
	\newcommand{\eeqn}{\end{eqnarray*}}

\makeatletter
\newcommand\figcaption{\def\@captype{figure}\caption}
\newcommand\tabcaption{\def\@captype{table}\caption}
\makeatother

\definecolor{shadecolor}{RGB}{220,220,220}

\tikzstyle{mybox} = [draw=black, fill=black!5, thick,
rectangle, rounded corners, inner sep=0pt, inner ysep=2pt]
\tikzstyle{fancytitle} =[fill=black, text=white]

\newcommand{\olist}{\texttt{Olist}\xspace}
\newcommand{\movie}{\texttt{MovieLens}\xspace}
\newcommand{\loyal}{\texttt{Loyal}\xspace}
\newcommand{\ped}{\texttt{PED}\xspace}
\newcommand{\event}{\texttt{Event}\xspace}
\newcommand{\eventnot}{\texttt{Event-Not}\xspace}

\newcommand{\fone}{\texttt{F1}\xspace}
\newcommand{\imdb}{\texttt{IMDB}\xspace}
\newcommand{\restbase}{\texttt{Restbase}\xspace}
\newcommand{\bio}{\texttt{Bio}\xspace}

\newcommand{\noresult}{\texttt{-}}

\newcommand{\base}{\texttt{Base}\xspace}
\newcommand{\bigtable}{\texttt{All}\xspace}
\newcommand{\randomselection}{\texttt{Random}\xspace}
\newcommand{\mutualinfor}{\texttt{MI}\xspace}
\newcommand{\backward}{\texttt{BE}\xspace}
\newcommand{\xgboost}{\texttt{XGBoost}\xspace}
\newcommand{\lightgbm}{\texttt{LightGBM}\xspace}
\newcommand{\randomforest}{\texttt{RF}\xspace}
\newcommand{\arda}{\texttt{ARDA}\xspace}
\newcommand{\ardano}{\texttt{ARDA-NoText}\xspace}
\newcommand{\leva}{\texttt{LEVA}\xspace}

\newcommand{\roc}{\texttt{AUC-ROC}\xspace}
\newcommand{\fscore}{\texttt{F1 Score}\xspace}
\newcommand{\accuracy}{\texttt{Accuracy}\xspace}
\newcommand{\average}{\texttt{Average Precision}\xspace}

\newcommand{\mae}{\texttt{MAE}\xspace}
\newcommand{\mse}{\texttt{MSE}\xspace}

\newcommand{\filterp}{\ell}

\newcommand{\vspacelen}{5}
\newcommand{\romannum}[1]{\uppercase\expandafter{\romannumeral #1 }}

\newcommand\vldbdoi{XX.XX/XXX.XX}

\newcommand\vldbavailabilityurl{URL_TO_YOUR_ARTIFACTS}

\begin{document}

\title{Graph-Based Feature Augmentation for Predictive Tasks on Relational Datasets}
\author{Lianpeng Qiao}
\email{qiaolp@bit.edu.cn}
\affiliation{%
    \institution{Beijing Institute of Technology}
}

\author{Ziqi Cao}
\email{3120235211@bit.edu.cn}
\affiliation{%
    \institution{Beijing Institute of Technology}
}

\author{Kaiyu Feng}
\email{kaiyufeng@outlook.com}
\affiliation{%
    \institution{Beijing Institute of Technology}
}

\author{Ye Yuan}
\email{yuan-ye@bit.edu.cn}
\affiliation{%
    \institution{Beijing Institute of Technology}
}

\author{Guoren Wang}
\email{wanggr@bit.edu.cn}
\affiliation{%
    \institution{Beijing Institute of Technology}
}







\begin{abstract}

Data has become a foundational asset driving innovation across domains such as finance, healthcare, and e-commerce. In these areas, predictive modeling over relational tables is commonly employed, with increasing emphasis on reducing manual effort through automated machine learning (AutoML) techniques. This raises an interesting question: can feature augmentation itself be automated and identify and utilize task-related relational signals?

To address this challenge, we propose an end-to-end automated feature augmentation framework, \sys, which enhances initial datasets using features extracted from multiple relational tables to support predictive tasks. \sys first captures semantic dependencies within each table by modeling intra-table attribute relationships, enabling it to partition tables into structured, semantically coherent segments. It then constructs a heterogeneous weighted graph that represents inter-row relationships across all segments. Finally, \sys leverages message-passing graph neural networks to propagate information through the graph, guiding feature selection and augmenting the original dataset. Extensive experiments conducted on ten real-life and synthetic datasets demonstrate that \sys consistently outperforms existing methods on both classification and regression tasks.

\end{abstract}

\maketitle



\section{Introduction} 
\label{sec:intro}

Across domains such as finance \cite{clements2020sequential}, healthcare \cite{van2010healthcare}, and the internet \cite{zhang2019deep}, relational table have emerged as a fundamental format for organizing structured data. To unlock the value embedded in such data, a broad range of predictive modeling tasks have arisen around relational tables. In response, the machine learning community has seen rapid progress in AutoML systems, which aim to automate the process of model selection and hyperparameter tuning. Provided with a dataset and task specification, these systems strive to achieve high performance with limited human involvement.

The effectiveness of the AutoML is greatly influenced by the quality of the user-provided raw data, particularly when the raw data features are insufficient. For instance, an event coordinator seeking to predict attendance for an upcoming event may find that historical data from prior, similar events could be beneficial due to its availability in databases. However, dependence solely on such historical data could result in imprecise predictions. This is attributable to multiple other determinants that could impact attendance numbers, including the event's geographical setting, the long-term residential trends of expected participants, their enthusiasm for the event's subject matter, and possibly their social connections, which might sway their decision to attend.

While databases conventionally organize all relevant information into relational tables to facilitate straightforward access, determining which features are predictive in many contexts remains a significant challenge. These predictive features are often dispersed across various interconnected tables within the database. A brute-force approach might involve merging all the potentially relevant tables into a single, extensive table. However, this method is fraught with problems. Primarily, such a large table would encompass an excessive number of features, leading to substantial demands on computational resources and increased processing duration. Ideally, extracting only those features that are genuinely beneficial would be adequate. 
Nevertheless, selecting attributes from tables to feed into an AutoML system for optimal predictive performance is challenging due to the following reasons:

\begin{itemize} [topsep=1pt,leftmargin=0.5cm]
    \item Feature relevance challenges (Selective Dependency) : In extensive databases, many attributes might lack direct predictive relevance for the target attribute. Including these irrelevant features can introduce noise, masking significant patterns. Beyond irrelevance, some attributes affect the target indirectly through complex correlations. These hidden dependencies, often subtle or combinatorial across various attributes, require in-depth domain knowledge or advanced analytical methods to detect.
    \item Table relationship challenges (Complex Dependency): Effectively leveraging information from multiple related tables introduces another layer of complexity. While the explicitly defined primary key-foreign key relationships offer a primary mechanism for data integration, the task becomes significantly more intricate when relational schema exhibit complex structural patterns. Understanding and navigating these complex inter-table relationships is crucial for building accurate predictive models.
\end{itemize}

In the face of these challenges, several research efforts have emerged. One major line of work focuses on 
selecting appropriate features for the predictive task from the merged single table. Classical methods such as Forward Selection and Backward Elimination   incrementally add or remove features to optimize model performance. However, they are prone to local optima. Forward Selection may overlook weakly correlated yet useful features, while Backward Elimination may mistakenly remove informative ones in the presence of noise \cite{chepurko2020arda}. Methods such as \cite{kumar2016join, shah2017key} focus on efficiency by avoiding unnecessary joins, particularly when foreign keys provide most required data. Nevertheless, these strategies are inherently cautious and generally do not enhance performance. ARDA \cite{chepurko2020arda} utilizes data discovery mechanisms to rank potential tables and employs heuristic methods for selecting features. As these rankings do not consider model specifics, their dependability is constrained. Additionally, ARDA is limited to single-hop joins, hindering its ability to leverage intricate multi-hop connections. In contrast, AutoFeature \cite{liu2022feature} and METAM \cite{galhotra2023metam} conceptualize augmentation as a sequential decision-making task, employing reinforcement learning models such as multi-armed bandits and deep Q-networks to dynamically explore join paths, thereby enhancing downstream performance. Nonetheless, these approaches typically entail expensive join operations and repetitive model assessments across diverse tables. Leva \cite{zhao2022leva}, while addressing inter-element relationships in relational databases, presupposes that the importance levels of adjacent elements are constant and not amenable to adaptive learning.

Recent approaches have explored transforming databases into graph structures capable of naturally representing intricate relationships between tables and efficiently managing multihop join paths. The methods proposed by \cite{cvitkovic2020supervised} and \cite{fey202relbench} involve mapping each table tuple to a graph node and employing primary key-foreign key (PK-FK) relationships as graph edges. However, this graph construction technique presumes that all features of the entire table are viewed collectively, neglecting the interaction relationships between individual features.


In this study, we present an innovative end-to-end automated framework named \sys. \sys serves as a tool that enhances initial datasets with features extracted from various relational tables, facilitating predictive tasks. The framework operates through three primary phases: Initially, it examines the associations between attributes across all tables (auxiliary tables) that can be joined or indirectly related to the initial data (base table) to discern semantic dependencies. This involves identifying attribute clusters, thereby partitioning the auxiliary tables into segments that are structurally and thematically refined. Secondly, using the segments derived from all auxiliary tables along with the fundamental data, a weighted heterogeneous graph model is constructed. This model characterizes the relational network of records in different table subsets. Lastly, the protocol used is the message passing mechanism of Graph Neural Networks (GNN), which facilitates the dissemination and aggregation of information within the graph structure. This mechanism is used to filter the attributes significant for feature augmentation using learning edge weights, thus enhancing the feature set of the initial data to improve the predictive performance. downstream.

Unlike table-based methods such as \cite{kumar2016join, shah2017key, chepurko2020arda}, the \sys framework provides specific advantages. (\romannum{1}) The utilization of a graph structure enables the depiction of diverse entities and their interactions via nodes and edges, resulting in a clear and concise representation of one-to-many, nested, and hierarchical relationships. This method significantly diminishes the complexity associated with join operations found in traditional systems, thus facilitating a more lucid and effective visualization of relationships. (\romannum{2}) The graph-based approach efficiently addresses the problem of data redundancy and inflation typical in conventional techniques. Unlike traditional strategies that require the transformation of all features into a high-dimensional matrix, graph approaches necessitate only the storage of nodes and their interconnections.

In contrast to other graph-based methodologies, the \sys framework presents distinct advantages by facilitating detailed feature enhancement via an analysis of attribute associations among tables and the identification of attribute groups, which results in more detailed thematic segmentation. It builds a weighted heterogeneous graph model that represents complex relationship networks through advanced weighting and heterogeneity techniques, surpassing basic join operations to improve the robustness and precision of node embeddings. Furthermore, \sys employs GNN to establish a robust information propagation mechanism, wherein the learning and optimization of edge weights aid in noise filtration and the selection of significant attributes for augmenting features.

\stitle{Contributions.} In summary, the contributions of this paper are:
\begin{itemize} [topsep=1pt,leftmargin=0.5cm]
    \item  We propose \sys, an automated feature enrichment framework that leverages GNNs over relational data graphs to improve prediction tasks with minimal manual effort. 
    \item In parsing attribute associations across tables, \sys identifies attribute groups, splitting the original tables into more refined thematic subsets, allowing the model to more accurately capture semantic dependencies within the data.
    \item We conducted extensive empirical research using multiple real-world and virtual relational datasets to evaluate our proposed method. The experimental results strongly validated the effectiveness and practical value of the \sys framework.
\end{itemize}


\section{Problem Formulation}
\label{sec:Preliminary}



\stitle{Preliminary. } Initially, we establish and clarify the essential concepts utilized across this paper.
\begin{itemize} [topsep=1pt,leftmargin=0.5cm]
    \item \textbf{Base Table} : The base table, denoted as $T_0 = \{a_1, a_2, ..., a_n, T \}$, is a unique table that stores the initial data without any augmentation.  Each $ a_i  ( i \in [1, n] )$ represents a feature attribute and $T$ denotes the target attribute.
    \item \textbf{Auxiliary Tables} : The auxiliary tables $\{T_1, T_2,\dots, T_K\}$ are a set of tables that could join the base table either directly or indirectly through the other auxiliary table. Each auxiliary table $T_k = {a^{k}_1, a^{k}_2, ..., a^{k}_n}$, where each $a^{k}_i(i \in [i,n])$ represents an attribute within the table $T_k$.
\end{itemize}

In a relational table, each tuple represents an entity or relationship instance, with attributes assigned to specific data types (e.g., numerical, string, or categorical). Tables are interconnected through keys: primary keys uniquely identify tuples, and foreign keys reference primary keys to enforce referential integrity.

\begin{figure}
    \centering
    \includegraphics[width=0.45\textwidth]{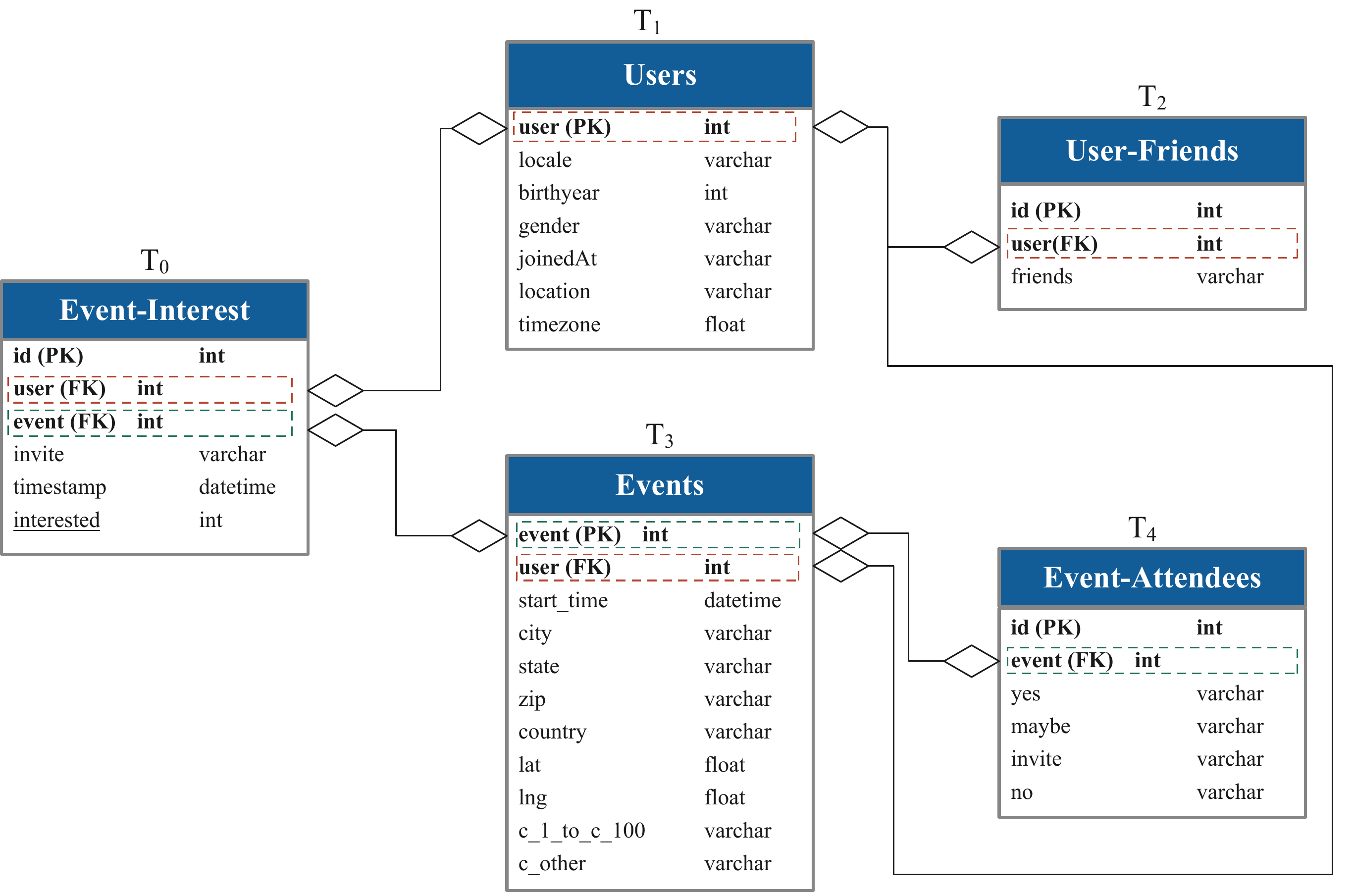}
    \caption{Event-Recommendation Relational Tables}
    \label{fig:join}
\end{figure}


Predictive tasks on relational tables learn mappings from input attributes to a target attribute. 
For instance, as shown in Figure ~\ref{fig:join} the \underline{interest} attribute in \underline{Event-Interest} is the target attribute, i.e., the recommendation system wishes to forecast the event of interest for each user. $T_0$ is the base table, while $T_1, \dots, T_4$ are auxiliary tables.
These tasks are categorized as classification (categorical target) or regression (numerical target). For example, predicting the status of a house sale is a classification task, 
whereas estimating its price is a regression.

Our objective is to predict user interest in specific events. In the \textit{Event-Attendees} (T4), the \textit{invited} and \textit{yes} attributes represent invitation status and confirmed attendance respectively. These indicators reflect event popularity and may influence user interest. Such attributes can correlate with our prediction target, providing valuable auxiliary signals. Identifying attribute correlations across auxiliary tables enhances predictive performance. However, detecting these relationships in complex database structures presents challenges. While some attributes are independently predictive, others gain value only in combination. Feature interactions reveal patterns invisible when examining features in isolation, requiring combinatorial analysis. For example, in \textit{Events} (T3), the \textit{lat} and \textit{lng} attributes individually offer limited information but jointly determine event location - a factor likely affecting interest. These combinations often embody complex semantic relationships that are difficult to manually define or encode. Such challenges motivate the development of an end-to-end learning framework capable of autonomously discovering synergistic attribute combinations. By enriching base table features through this process, we can improve model performance while minimizing manual feature engineering efforts.

\noindent\textbf{Problem Statement.} Given a base table $T_0$ with the target attribute and auxiliary tables $\{T_1, \dots, T_K\}$, we aim to augment $T_0$ with attribute from $\{T_k\}$ to improve prediction accuracy for the target. However, the core challenge lies in identifying relevant attributes from $\{T_k\}$ whose integration into $T_0$ enhances model performance.



\section{Solution Overview} 
\label{sec:overview}

\begin{figure*}[t]
    \centering
    \includegraphics[width=1\textwidth]{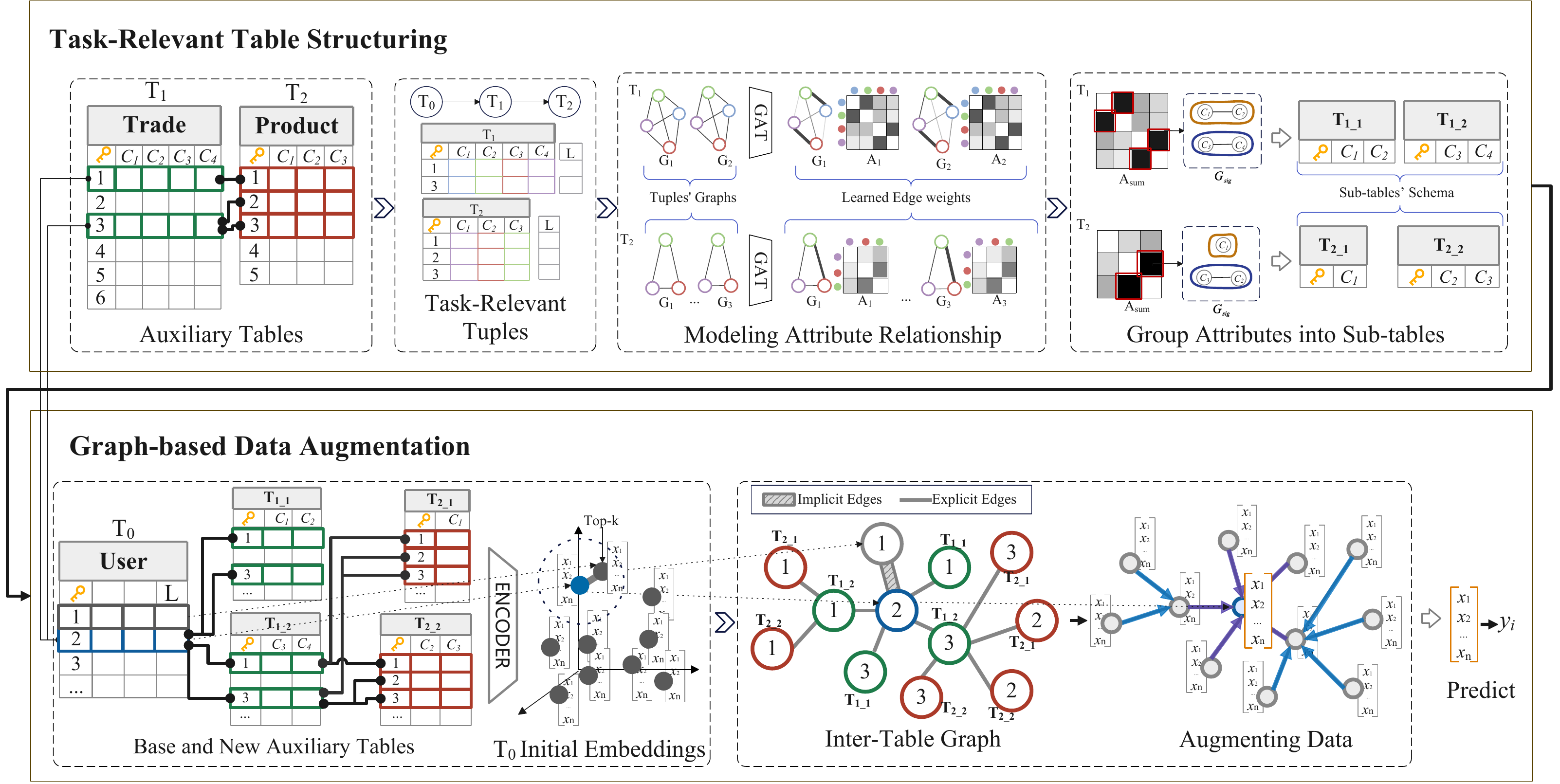}
    \caption{\text{\sys} Framework}
    \label{fig:framework}
\end{figure*}




As previously discussed, integrating discriminative features from auxiliary tables into the base table poses two key challenges: assessing feature importance and modeling inter-table relationships. We propose \sys, a comprehensive framework (illustrated in Figure ~\ref{fig:framework}) to address these challenges. In the first phase, \sys uses a graph neural network to identify task-relevant links among table attributes, segmenting the original tables into semantically aligned sub-tables. In the second phase, \sys forms a heterogeneous graph incorporating both explicit inter-table connections and implicit similarity ties. A graph neural network is employed to refine the tuple representations in the base table, thereby improving features for the predictive task.

\stitle{Stage One. }In the first stage of \sys, called "Task-Relevant Table Structuring," our goal is to extract from each original table a set of sub-tables composed of highly related (and potentially overlapping) attributes. This approach moves away from the traditional practice of treating a table as a monolithic structure, instead generating semantically coherent attribute groups that are more directly aligned with the prediction task. Intuitively, not all tuples are useful for the task, as many cannot be directly or indirectly joined with the base table, rendering them uninformative for prediction. Therefore, we first identify the most task-relevant tuples in each table based on their potential contribution to the prediction objective. By focusing subsequent analysis on this informative subset, we enhance both the efficiency and effectiveness of feature modeling and downstream learning.

\stitle{Stage Two. }In the second stage of \sys, we aim to enhance the feature representation of the base table by integrating information from the sub-tables generated in the first stage. Intuitively, not all sub-tables contribute equally to the prediction task. Thus, the key challenge at this stage is to select the most task-relevant sub-tables and effectively integrate them into the base table. Moreover, since it is difficult to precisely evaluate whether the intra-table relationships identified in the first stage fully align with the prediction task requirements, we assess the effectiveness of feature enhancement based on the final prediction performance in the second stage. Therefore, we design a heterogeneous graph and employ a Graph Neural Network (GNN) to learn edge weights, thereby identifying and filtering attributes that are meaningful for feature enhancement. While achieving this objective, the GNN also aggregates relevant information from associated sub-tables to the base table nodes, ultimately generating highly valuable representations for downstream prediction tasks. The learned representations can be directly applied to different prediction heads (e.g., classifiers or regressors), supporting diverse downstream applications.
\section{Task-Relevant Table Structuring} 
\label{sec:column_relaitonship}

To enhance the predictive performance of the base table, we aim to incorporate useful features from auxiliary tables in the database. Constructing effective attribute sets from auxiliary tables is non-trivial, as individual attributes may not be discriminative alone but become useful in combination, while the tables themselves contain heterogeneous data types.To tackle these obstacles, we introduce a method for modeling intra-table attribute relationships within each supplementary table. Each tuple is viewed as an instance of attribute co-occurrence, enabling the creation of a fully connected graph among attributes. A Graph Attention Network (GAT) is then employed to learn the interaction strengths. Task-relevant attribute combinations are identified through learned edge weights and used to form more focused, informative sub-tables. The methodology involves three steps: (1) pinpointing task-relevant tuples for base table integration to gain supervision; (2) constructing attribute-level graphs for these tuples and training a GAT to capture feature interactions; and (3) deriving highly relevant attribute groups from attention weights to reorganize each supplementary table into a compact, task-specific sub-table.

\subsection{Identifying Task-Relevant Tuples}
Recall that our goal is to uncover meaningful combinations of attributes within each table that contribute to predicting the target attribute's value. To achieve this, we analyze the relationships between attribute values within individual tuples of the table. Identifying task-relevant tuples is non-trivial. The natural intuition suggests that a tuple is task-relevant if it can be linked to a target attribute. However, given the complex schema relationships, a single tuple may connect to multiple target values. So the key question to be answered is: Which tuples are task-relevant and useful for learning attribute relationships?

Intuitively, a tuple in an auxiliary table is considered potentially relevant to a target attribute value if it can be linked to a tuple in the base table. This linkage is established through a sequence of join operations defined by the relational schema. However, relational schema can be complex, offering numerous join paths. Many of such join paths may not capture semantically meaningful relationships for our specific prediction task. To address this, we construct meta-path for each auxiliary table to define its semantic relationship to the base table. 
As described earlier, the join relationships between the base table 
$T_0$ and all auxiliary tables $\{T_1, ..., T_K\}$. are defined by the database schema. This relationship can be effectively modeled as a Directed Join Graph (DJG), wherein each node represents a relational table within the dataset $T = \{T_0\} \cup \{T_1, ..., T_K\}$, while the edges signify executable join operations in $T$, incorporating their respective link types such as $1:1$, $ 1:n$, and $n:1$. However, an unprocessed DJG presents complex pathways, which complicates the handling of join relationships, particularly those involving circular structures. To effectively identify a meaningful meta-path for each auxiliary table, we employ a greedy path search algorithm starting from the base table $T_0$, expanding the path one step at a time. At each step, the algorithm chooses the next join that leads to the highest path score. The path scoring function jointly considers two aspects:

\begin{itemize} [topsep=1pt,leftmargin=0.5cm]
    \item \textbf{Path length:} Shorter paths are generally more interpretable and less likely to introduce redundant or noisy features. The path length score is defined as $S_L = \frac{1}{1 + L}$,
    where $L$ is the number of edges in the current path.
    
    \item \textbf{Join directionality:} One-to-many (1:n) joins are prone to amplifying noise and redundancy, especially when they appear repeatedly in multi-hop paths. To model this, we define a weighted penalty term that accounts for the risk associated with each 1:n join:
    
    $S_N = \frac{1}{1 + \sum_{i=1}^{L} w_i \cdot \mathbb{I}_{\text{1:n}}(e_i)}$, where $e_i$ denotes the $i$-th edge in the path, $\mathbb{I}_{\text{1:n}}(e_i)$ is an indicator function that equals 1 if the edge is a 1:n join, and $w_i$ is a weight reflecting the risk (e.g., average fan-out) of the join.
\end{itemize}

The overall path score is computed as a weighted sum: $S_{\text{path}} = \alpha S_L + \beta S_N$
where $\alpha$ and $\beta$ are tunable weights that balance structural simplicity and join stability. Other domain knowledge or schema-level constraints can also be incorporated into this scoring function.

During greedy search, the algorithm expands only the neighbor with the highest incremental path score at each step, avoiding the cost of exhaustive enumeration. Once the auxiliary table is reached, the path with the highest accumulated score is selected as its meta-path.

\begin{figure}
    \centering
    \includegraphics[width=0.45\textwidth]{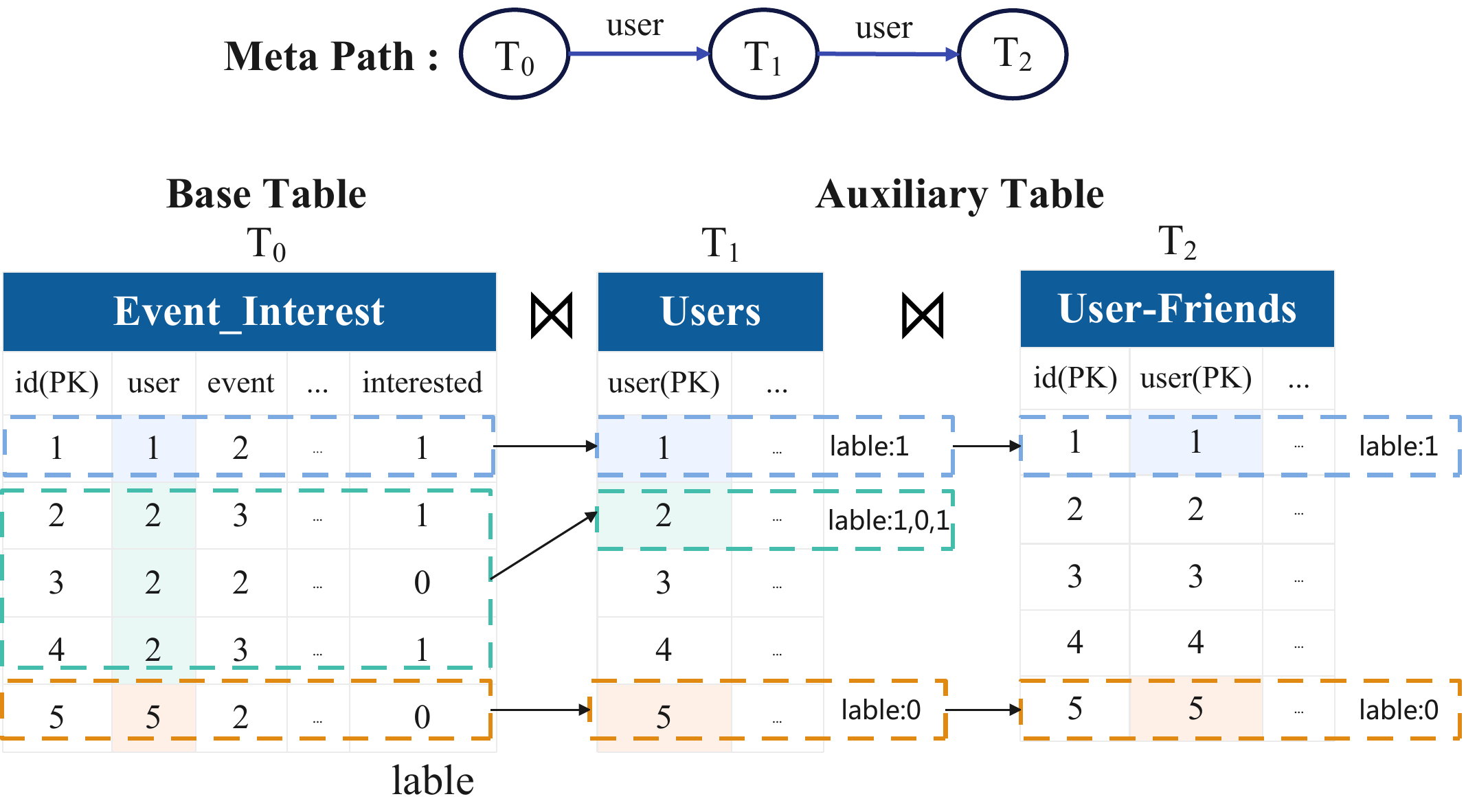}
    \caption{An Example of Join Instance}
    \label{fig:join-example}
\end{figure}

Given an auxiliary table $T_i=\{t_{i1}, \dots, t_{in}\}$ and its specified meta-path, if there exists an instance of a join path that successfully links a tuple $t_{ij}$ in $T_i$ to a tuple $t_{0k}$ in the base table, we consider tuple $t_{ij}$ relevant to tuple $t_{0k}$ and thus carrying the corresponding target label. Consider the example in Figure~\ref{fig:join-example}. As shown in the figure, the base table is \textit{Event-Interest}, with the target attribute interest and \textit{Users} serving as an auxiliary table. The tuple with the primary key $user = 1$ in the \textit{User} table obtains the value of the label from the Base table by linking to the tuple in the \textit{Event-interest} table where the foreign key $user = 4$. 



\stitle{Coreset.} In real-world scenarios, database tables can contain a vast number of tuples. Utilizing every tuple ensures information completeness but results in extremely high computational costs for further processing. To mitigate this, we employ a sampling strategy: forming a coreset by selecting a representative subset of tuples from each auxiliary table to approximate the entire dataset. This coreset maintains essential structural details while significantly lowering computational costs, albeit with a slight potential compromise in model accuracy. Our approach initiates coreset construction from the base table, selecting core tuples based on criteria such as label distribution. These core tuples are then extended to auxiliary tables along meta-paths.

\subsection{Modeling Attribute Relationship with GAT}
Building upon the task-relevant tuples identified in Section 4.1, we now turn our attention to discerning the predictive power of individual attributes within these auxiliary tables. However, it is non-trivial task, as the contribution of each attribute to the target prediction task is not equal. Some might be directly indicative of the target, while others might be noisy or only relevant in conjunction with other attributes. To effectively capture these complex inter-attribute dependencies and automatically learn their relevance to the prediction task, we propose to represent each task-relevant tuple as a complete graph. 

\subsubsection{Construct a Complete Graph for Each Tuple.}
For each identified task-relevant tuple in an auxiliary table, we denote its non-key attributes as $\{c_1, c_2, \dots, c_N\}$. Here we explicitly exclude the primary key of the auxiliary table itself and any foreign keys it might contain, as their inherent values primarily serve for identification and inter-table linkage rather than directly representing features relevant to the prediction task within the scope of this table's attributes. For such tuple $t$, we construct an individual complete graph $G_t=( \mathbf{V}_t,  \mathbf{E}_t,  \mathbf{X}_t)$. The set of nodes in $V_t$ corresponds to the attribute values in $\{c_1, c_2, \dots, c_N\}$ within the tuple $t$. Given the heterogeneous data types of the attribute values (numerical, categorical, and text), the feature vector $ \mathbf{X}_t[i]$ for node $v_i$ is encoded according to the data type of the corresponding attribute as follows:




\stitle{Numerical}: For all task-relevant tuples, identify the data type by column. For each element $c_i$ of the identified numeric type columns (from any tuple $t$), use a model\cite{gorishniy2022embeddings} with shared parameters to encode it to obtain the encoded result $\mathbf{e}_i \in \mathbb{R}^{d_{\text{num}}}$.


\stitle{Categorical}: For each cell $c_i$ recognized as belonging to a categorical column, we employ an encoding mechanism utilizing an embedding look-up based encoder to transform it into a vector $\mathbf{e}_i \in \mathbb{R}^{d_{\text{cat}}}$.

\stitle{Text}: Regarding text columns, there are several impressive models available, such as Sentence-BERT \cite{reimers2019sentence}. Encode the text category $c_i$ into a vector $\mathbf{e}_i \in \mathbb{R}^{d_{\text{text}}}$.

For each cell $ c_i $, generate embeddings based on its data type:
\begin{equation}~\label{eqa:encode}
    \mathbf{e}_i = 
    \begin{cases} 
    f_{\text{num}}(c_i) \in \mathbb{R}^{d_{\text{num}}}, & \text{if } \text{type}(c_i) = \text{num} \\
    f_{\text{cat}}(c_i) \in \mathbb{R}^{ d_{\text{cat}}}, & \text{if } \text{type}(c_i) = \text{cat} \\
    f_{\text{text}}(c_i)\in \mathbb{R}^{ d_{\text{text}}}, & \text{if } \text{type}(c_i) = \text{text}
    \end{cases}
\end{equation}
where $ f_{\text{num}} $, $ f_{\text{cat}} $ and $ f_{\text{text}} $ are the numerical encoder, the categorical encoder, and the text encoder with output dimensions $ d_{\text{num}} $, $ d_{\text{cat}} $, and $ d_{\text{text}} $, respectively. In the experiment, we use PyTorch Frame, an advanced modular deep learning extension for PyTorch \cite{hu2024pytorch}.

Embedding dimensions generated by encoders for different cell categories (e.g., numerical, categorical, text) often exhibit significant inconsistencies. Specifically, text-type cells typically yield higher-dimensional embeddings compared to their numerical and categorical counterparts. To address this heterogeneity, deep learning methods such as Multi-Head Self-Attention mechanisms \cite{vaswani2017attention} or adapted ResNet \cite{gorishniy2021revisiting} architectures can be employed to project column-wise embeddings into a unified dimensional space, thereby enabling seamless integration for downstream tasks.

Project embeddings of varying dimensions into a unified target dimension $ d_{\text{out}} $:
\begin{equation}~\label{eqa:representation}
    \mathbf{h}_j = \mathbf{e}_{j} \mathbf{W} \in  \mathbb{R}^{d_{out}}
\end{equation}
where $\mathbf{W}$ are learnable projection matrices for dimensionality alignment to $ d_{\text{out}} $.

\begin{figure}
    \centering
    \includegraphics[width=0.4\textwidth]{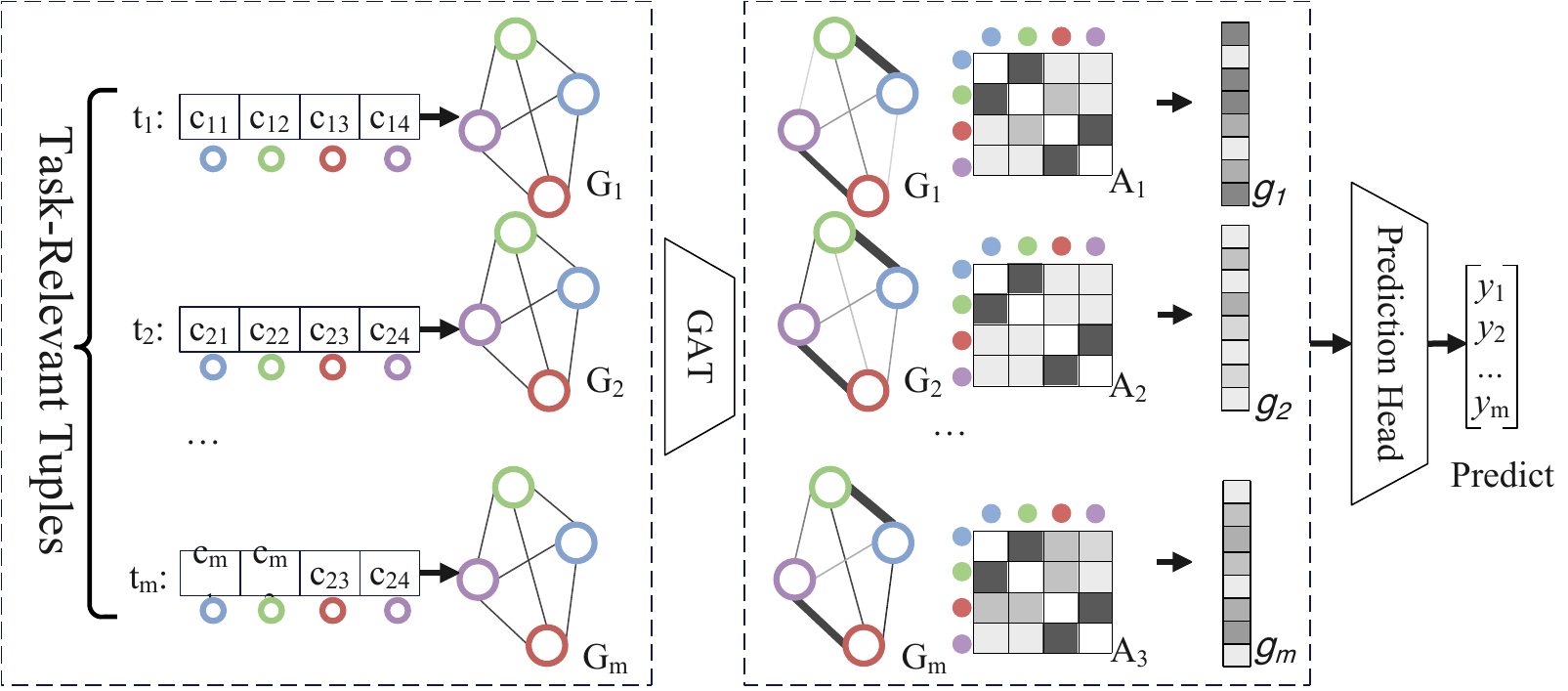}
    \caption{Modeling Attribute Relationship with GAT}
    \label{fig:gat}
\end{figure}

\subsubsection{Leveraging GAT for Attribute Relationship}
Having constructed a complete graph for each task-relevant tuple, where nodes represent attributes and edges signify potential relationships, the next crucial step is to capture the intricate and task-specific relationships between attributes within each tuple. The general idea is shown in Figure~\ref{fig:gat}. To achieve this, we first build a complete graph for each identified task-relevant tuple. These complete graphs are encoded by a share-parameter GAT. The representation of each graph is fed into a prediction head, aiming to predict the label associated with the corresponding tuple. Unlike methods that treat all nodes uniformly, GATs employ an attention mechanism that allows each node to dynamically learn the importance of its connection with other nodes in the graph in the context of the target prediction task. The learned attention weights on the edges of the complete graphs serve as indicators of the significance of these pairwise node interactions (representing attribute relationships) for predicting the target attribute label. 


Given the complete graph $G_i(V, E, X_i)$ constructed from the $i$-th task-relevant tuple, let $u$ and $v$ be two nodes in the graph. The attention weight $ \alpha^i_{uv} $ is derived as follows:  
\begin{equation}~\label{eqa:gat}
    \begin{gathered}
     \mathbf{h}^i_u = \mathbf{W} \mathbf{X}^i[u] , \quad \mathbf{W} \mathbf{h}_v = \mathbf{X}^i[v] , \\
     e^i_{uv} = \text{LeakyReLU}\left( \alpha ^\top \left[ \mathbf{h}^i_u \mathbin{\|} \mathbf{h}^i_v \right] \right), \\
    \alpha^i_{uv} = \frac{\exp(e^i_{uv})}{\sum_{w \in V} \exp(e^i_{uw})}
    \end{gathered} 
\end{equation}

where $X^i[u]$ and $X^i[v]$ are feature vectors for $u$ and $v$. The attention mechanism $\alpha$ and weight matrix $W$ of the GAT layer are shared across all complete graphs. The updated embedding $h_{v}^{i}\,'$ for node $v$ is computed by aggregating the feature vectors of its neighbors, weighted by the learned attention coefficient
\begin{equation}
     \mathbf{h}^i_v\,'=\sigma(\sum_{u}\alpha^i_{uv} \mathbf{W}\,'  \mathbf{h}^i_u)
\end{equation}
where $u$ is any other node in the complete graph, $h_u$ is the embedding of $u$, $W'$ is a learnable weight matrix, and $\sigma$ is a non-linear activation function. 

The graph-level representation $ \mathbf{g}$ of the complete graph $G_i$ is derived by applying a pooling function to the set of its learned node embeddings $\{ \mathbf{h}^i_v\}_{v\in V}$ as follows:
\begin{equation}
     \mathbf{g}^i = \text{POOL}( \mathbf{h}^i_v | v \in V)
\end{equation}

The resulting graph-level representation $g^i$, which encapsulates the learned relationships between attributes within the corresponding task-relevant tuple, is fed into a prediction head. The architecture of this prediction head is specifically tailored to the type of the target attribute in the base table. The specific architecture of the prediction head is orthogonal to our approach. Any standard network suitable for the target task (classification or regression) can be employed. For a regression task, the prediction head outputs a continuous value $\hat{y_i}$, and the loss $\mathcal{L}_i$ is computed using Mean Squared Error (MSE) between $\hat{y_i}$ and the true target attribute value $y_i$. In contrast, if it is a classification task, the prediction head outputs a probability distribution over the possible classes $\hat{p}_i$, and the loss $\mathcal{L}_i$ is computed using cross entropy loss between $\hat{p}_i$ and the one-hot true class label $y_i$. The total loss is computed by aggregating the individual prediction losses as follows:
$$\mathcal{L}_{total}=\sum_{i=1}^{B}\mathcal{L}_i$$


The GAT mechansim learns attention weights for every edge in each complete graph. These learned weights are derived directly during the optimization process aimed at minimizing the prediction loss. Thus, they provide a valuable insight into the contribution of each attribute pair to the final prediction. Higher weights signify more influential attribute relationships for the predictive task. These learned attribute relationships will be leveraged to group attributes into more focused sub-tables.


\subsection{Group Attributes into Sub-tables}
Having modeled the pairwise relationships between attributes within each task-relevant tuple using GAT and obtained edge weights reflecting their predictive significance, we now aim to leverage this information to structure the original auxiliary tables into more focused, task-relevant sub-tables. To begin with, we first introduce how to extract significant attribute relationships based on the edge weights learned from all task-relevant tuples. 

\subsubsection{Extracting Significant Attribute Relations.}
For each task-relevant tuple $t_i$, we use $\mathbf{A}_i$ to denote the learned edge weights, where each entry $\mathbf{A}_i[u][v]=\alpha^i_{u,v}$, the attention weight between $u$ and $v$ in graph $g^i$. To consolidate the importance of these pairwise relationships across all task-relevant tuples, we compute a cumulative attribute relationship matrix $\mathbf{A}_{sum}$ by summing the individual weight matrices:
\begin{equation}
    \mathbf{A}_{sum} = \sum_{i=0}^{B}\mathbf{A}_i
\end{equation}
To ensure a consistent scale for identifying significant relationships, we then normalize the values in $\mathbf{A}_{sum}$ to the range $[0,1]$ using min-max normalization. This normalized matrix $\mathbf{A}_{norm}$ reflects the overall importance of each pairwise attribute relationship across the entire set of task-relevant tuples. We then identify the most important attribute relationships by applying a predefined threshold $\filterp$. An edge between node $u$ and $v$ is considered significant if its corresponding weight $\mathbf{A}_{norm}[u][v]$ is greater than $\filterp$. We use $E_{sig}$ to denote the set of selected edges. These selected edges represent the attribute pairs that the model has consistently deemed most relevant for predicting the target variable across all considered contexts.

Identifying significant pairwise attribute relationships provides valuable insights, yet directly utilizing the original wide table for downstream tasks can still be suboptimal. A key challenge lies in the fact that not all identified relationships might be equally relevant in all contexts, and the presence of weakly related or irrelevant attributes within the same table can introduce noise and dilute the more potent signals. Therefore, organizing the attributes into more focused sub-tables, based on their learned relationships, holds the potential to create more semantically coherent and task-relevant data segments. 

\subsubsection{Extracting sub-tables}
Motivated by the potential benefits of focusing on strongly related attributes, we now introduce our strategy for extracting more targeted sub-tables from the original auxiliary table. To achieve this extraction, we construct a graph $G_{sig}( \mathbf{V},  \mathbf{E}_{sig})$ to explicitly model the relationship between attributes based on the set $ \mathbf{E}_{sig}$ of selected significant edges. By analyzing the structure of this graph formed by gathering all significant edges, we aim to discover tightly knit groups of highly relevant attributes. Specifically, we propose to identify the set of maximal cliques $\{C_1, C_2, \dots, C_t\}$ within $G_{sig}$. Let be the set of identified maximal cliques. For each maximal clique $C_i$, the attributes corresponding to the nodes within it form a sub-table. The rationale behind using maximal cliques is that each clique represents a subset of attributes where every attribute is significantly related to every other attribute within that subset, thus forming a highly cohesive and potentially task-relevant sub-table. 
\begin{figure}
    \centering
    \includegraphics[width=0.4\textwidth]{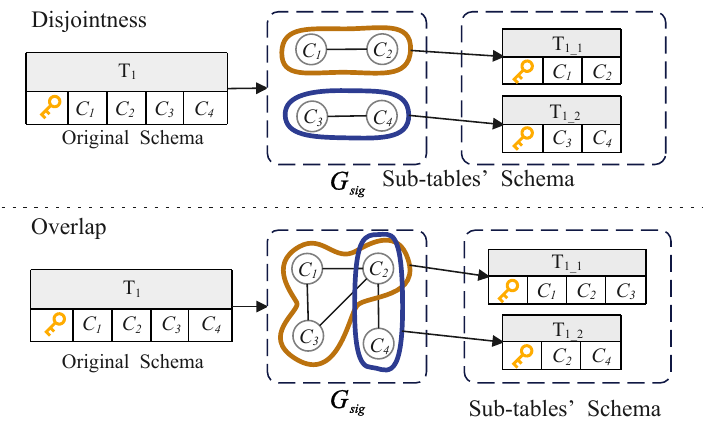}
    \caption{Scenarios for Extracting sub-tables from a Table}
    \label{fig:feature}
\end{figure}
\begin{example}
    As shown in Figure 6, in the overlap scenario, the attributes of the auxiliary table $T_1$ form the graph $G_{sig}$. In this graph, the nodes $C_1$, $C_2$, and $C_3$ are all interconnected, forming a maximal clique $\{C_1, C_2, C_3\}$, which constitutes a sub-table $T_{1,2}$.
\end{example}

\stitle{Remark.}
It is important to note that this strategy may result in sub-tables that are either disjoint or overlapping. As shown in Figure~\ref{fig:feature}, \textit{Disjoint} produces two disjoint sub-tables, while \textit{Overlap} produces two overlapping sub-tables. This potential for disjointness and overlap reflects the complex and potentially multifaceted nature of attribute relationships relevant to the prediction task.

\section{Graph-based Data Augmentation} 
\label{sec:feature_augmentation}
Following the modeling of intra-table attribute relationships to capture semantic dependencies and partition tables into more focused segments, as detailed in the previous section, we now turn our attention to the crucial task of modeling inter-table relationships. In this section, we introduce a approach to construct a heterogeneous graph that explicitly encodes the connections between tuples across these partitioned tables and the base table. This enables us to select task-relevant attributes and augment init tuples that span the entire relational schema.

\subsection{Constructing the Inter-Table Graph}
While the partitioned sub-tables offer more focused representations of information within individual tables, the true predictive power often lies in the intricate relationships between entities residing across these different tables. To effectively capture such inter-table dependencies, a graph-based learning emerges as a natural and intuitive choice. However, the construction of such a graph in our context presents two \textbf{key challenges} that necessitate careful consideration: 
\begin{itemize} [topsep=1pt,leftmargin=0.5cm]
    \item \textbf{Original v.s. Splitted Tuples.} How do we effectively link base table tuples to relevant information distributed across the partitioned auxiliary tables? Moreover, how are inter-table relationships accurately modeled when a single original tuple is represented by multiple sub-tuples, ensuring proper information aggregation?
    \item  \textbf{Supervision Isolation.}Since only base table nodes are labeled and the graph structure is defined exclusively by join relationships, semantically similar nodes are often disconnected in the topology, which limits the propagation of supervision signals and hinders the learning of consistent representations.
\end{itemize}

To address these two key challenges, we propose a novel approach for constructing a heterogeneous graph.
The heterogeneous graph, denoted by $\mathcal{G} = (\mathcal{V}, \mathcal{E},  \mathcal{X})$, is constructed as follows:

\stitle{Nodes.} Each tuple in the base table $T_0$ and each tuple within the partitioned sub-tables $T_{i\_1}, \dots, T_{i\_j}$ of the auxiliary tables $T_i$ constitutes a node in $\mathcal{G}$. Thus, the set of nodes $\mathcal{V}$ is the union of all tuples from the base tables and all sub-tables from the partitioned auxiliary tables. 

\stitle{Features.} Each node in the heterogeneous graph, representing a tuple, is associated with a feature vector derived from its constituent attribute values. Given the diverse nature of attributes in relational tables, which can be numerical, categorical, or textual, we employ modality-specific encoders to generate meaningful embeddings for each attribute. Following the idea in \cite{fey202relbench}, we use a modality-specific encoder to embed each attribute into embeddings. Once each attribute within a tuple is embedded, we concatenate these attribute embeddings to form the final feature vector for the node.

\stitle{Edges.} We define two primary types of edges within our heterogeneous graph, capturing both implicit relationships within the base table and explicit, schema-defined connections across tables:

\begin{itemize} [topsep=1pt,leftmargin=0.5cm]
    \item Explicit Edges(Inter-Table Connections): Consider tuple $t_u$ from one table and  tuple $t_v$ from a different table. We add an edge between $t_u$ and $t_v$ if tuple $t_u$ and $t_v$ can be joined based on the primary and foreign key relationship defined in the database schema. Moreover, if an original auxiliary table $T_i$ is partitioned into sub-tables $T_{i1}, \dots, T_{im}$, a tuple $t$ from $T_i$ is correspondingly represented by sub-tuples $t_1\in T_{i1}$, $t_2\in T_{i2}$, $\dots$, $t_m \in T_{im}$. If a tuple $t'$ from another table can be joined with the original tuple $t$, we create edges from $t'$ to each of the sub-tuples of $t$, i.e., $(t', t_1), \dots, (t', t_m)$. This ensures that information from a joinable tuple is propagated to all the tuples from the partitioned sub-tables.
    \item Implicit Edges(within the Base Table): Tuples in the base table might represent entities that are semantically similar or related in ways not directly encoded by joins. For instance, two customer records might have very similar purchasing histories or demographic profiles, even if there isn't a direct foreign key relationship linking them. To capture such semantic similarity, we create edges between tuples in the base table based on the similarity between their corresponding feature vectors. Specifically, for any two tuples from the base table $T_0$, we create an edge between them if the similarity between their features is above a predefined threshold $\theta$. Alternatively, we can adopt a more selective approach by connecting each tuple to its top-$K$ most similar tuples within the base table. By limiting the number of implicit connections per node, we can create a sparser graph compared to a threshold-based approach. This can lead to more efficient message passing during GNN training and potentially reduce computational overhead.
\end{itemize}

\subsection{Augmenting Table}
The constructed heterogeneous graph serves as a comprehensive representation of the intricate relationships among tuples spanning the base table and all partitioned auxiliary tables. However, to effectively leverage this complex structure for both feature selection and feature enhancement, we require a mechanism capable of performing information propagation and aggregation over the graph, thereby learning fine-grained edge weights that capture these diverse relational patterns.


GNNs perform message passing to allow each node to gather information from its neighbors, weighted by the edge connections, and iteratively refine its own representation. This process enables the model to learn high-order relationships and dependencies that are implicitly encoded within the graph structure, ultimately yielding tuple embeddings that are informed by the broader relational context and tailored for accurate prediction of the target column in the base table.

Consider a heterogeneous graph $G = (\mathbf{V}, \mathbf{E}, \mathbf{T_v}, \mathbf{T_e})$, where $\mathbf{V}$ is the set of nodes, $\mathbf{E}$ is the set of edges, $\mathbf{T_v}$ is the set of node types, and $\mathbf{T_e}$ is the set of edge types. 

\stitle{Aggregation Function} 
For a target node $v_i$ from the base table and an edge type $t \in T_e$, messages are generated from its neighbors $N_{v_i}$ based on different edge types:
\begin{equation} ~\label{aggre}
    m_{i}^{t} = \sum_{j \in N_{v_i}^{t}} M(h_j, h_{e_{ji}}, h_i)
\end{equation}
Here, $M$ is the message generation function, $h_j$ is the feature of neighbor node $j$, $h_{e_{ji}}$ is the edge feature, $h_i$ is the feature of node $i$, and $N_{v_i}^{t}$ denotes the set of neighbor nodes connected to node $i$ via edges of type $t$.

\stitle{Message Passing} 
The messages $m_i^t$ generated from each edge type $t \in T_e$ are integrated:
\begin{equation}
    h_i^{(l+1)} = U(h_i^{(l)}, \{m_i^t \mid t \in T_e\})
\end{equation}
Here, $U$ is the message updating function, $h_i^{(l)}$ denotes the feature of node $i$ at layer $l$, $h_i^{(l+1)}$ denotes the feature of node $i$ at layer $l+1$, and $\{m_i^t \mid t \in T_e\}$ is the set of all messages received from different edge types.

\stitle{Prediction} The total loss is computed by aggregating the prediction losses over all labeled nodes:
\begin{equation}~\label{eqa:loss}
  \mathcal{L}_{total} = \sum_{i=1}^{N} \mathcal{L}(y_i, f_{G}(x_i))
\end{equation}
where $N$ is the number of labeled nodes, $x_i$ denotes the features of node $i$, $y_i$ is its ground-truth label, $\mathcal{L}$ is the task-specific loss function, and $f_{G}$ is a heterogeneous graph neural network.

Our research aims to use a GNN to develop advanced representations for tuples in the base table by integrating vital data from auxiliary tables. Given that labels exist solely for the base table nodes, we employ the GNN to spread label data, boosting the base table node representations' accuracy and predictive ability. Specifically, by refining the graph's edge weights, we evaluate each edge's significance to improve predictions, assuming more important edges receive higher weights during training. The central task is to boost the tuple representations' predictive accuracy from the base table. We test these enriched representations in real-world prediction contexts to pinpoint the most valuable features. This issue is defined as enhancing prediction efficacy according to the formula \ref{eqa:loss}. This strategy results in selecting features linked to high-weight edges, marking them as crucial for the prediction task. The enhanced representations from the base table not only reflect explicit inter-table relationships but also capture hidden semantic connections, thereby achieving superior predictive accuracy amidst intricate patterns across diverse tables.

\section{Experimental Study} 
\label{sec:experiment}

In this section, we focus on answering the following questions:

\noindent\textbf{RQ1:} How does \sys perform compared to various baseline?

\noindent\textbf{RQ2:} How do key parameters affect the performance of \sys?

\noindent\textbf{RQ3:} How do different components contribute to the system's performance?

\subsection{Experimental Settings}

\noindent \textbf{\textit{Datasets.}} We use 10 datasets from various domains to evaluate our \sys. As shown in Table \ref{tab:datasets}, we give the basic information on the datasets. Among the datasets, six of them are for classification tasks and the other four datasets for regression tasks. Now we present the detailed information of these datasets.

\begin{itemize} [topsep=1pt,leftmargin=0.5cm]
    \item \olist ~\cite{olist} is derived from the Brazilian e-commerce platform Olist and is utilized for predicting customer satisfaction. A customer is deemed satisfied if their review$\_$score is at least 3. 
    \item \movie ~\cite{movie} is compiled by GroupLens Research from the MovieLens platform aiming to forecast user gender.
    \item \loyal ~\cite{loyal} aims to enhance credit card holders' personalized recommendations by employing machine learning for predicting customer loyalty.
    \item \ped ~\cite{ped} from DonorsChoose.org is an online charity aiding K-12 schools through project-based donations. The task is to identify exceptionally exciting projects. 
    \item \event ~\cite{event} includes user actions, event metadata, and other information, in order to predict which events users will be interested in.
   \item \eventnot ~\cite{event}
    aligns with the event dataset but focuses on predicting the events that are not of user interest.
    \item \fone ~\cite{f1} contains comprehensive historical data from all Formula One seasons since 1950. The regression task involves predicting the average position order of a driver.
    \item \imdb ~\cite{imdb} predicts the scores of movies and the dataset includes individual characteristics of movies.
    \item \restbase ~\cite{restbase} comprises restaurant data, framing the prediction of customer review details as a regression task.
   \item \bio ~\cite{bio} dataset is utilized for a regression task aimed at estimating "the molecule's bioactivity".
\end{itemize}
\noindent  \textbf{\textit{Baseline.}} Our methodology is evaluated against various solution.
\begin{enumerate}[topsep=1pt,leftmargin=0.5cm]
    \item \textbf{\base} retains only the native features of the base table.
    \item \textbf{\bigtable} simply joins all auxiliary tables with the base table.
    \item  \textbf{\randomselection} randomly selects $K$ features from auxiliary tables.
    \item  \textbf{Mutual Information (\mutualinfor)} is the filter-based feature selection method that evaluates the importance of features through different mechanisms and selects the top-$K$ feature subset.
    \item \textbf{Backward Elimination (\backward)} \cite{guyon2003introduction} is a wrapper-based feature augmentation method that starts by joining all auxiliary tables and then iteratively removes the features that most degrade model performance. The final subset is achieved by step-by-step exclusion of the least useful features.
    \item  \textbf{\xgboost, \lightgbm and Random Forest (\randomforest)} \cite{chen2016xgboost, ke2017lightgbm, breiman2001randomf} are embedded feature selection methods that combine the advantages of filter methods and wrapper methods by automatically performing feature selection during the model training process.
    \item \textbf{\arda} \cite{chepurko2020arda} combines the methods of building the core set and selecting random injection features, effectively screening out features that contribute to improving model accuracy while avoiding the introduction of noisy features. As the \ardano algorithm is incapable of directly handling text-type data, we implement two approaches: either converting text attributes to high-dimensional vectors (\textbf{\arda}) or eliminating them entirely (\textbf{\ardano}).
    \item \textbf{\leva} \cite{zhao2022leva} is an end-to-end system that constructs relational embeddings to represent relational data as vectors, which are then used to efficiently describe the base table. It constructs a graph to represent relational data and embeds this graph into high-dimensional space.
\end{enumerate}

\begin{table}
\small
  \caption{Statistics of Datasets}
  \label{tab:datasets}
  \begin{tabular}{c|c|c|c|c}
  \toprule
    Dataset &Tables & Features&  Rows& Task \\
\hline
   \olist&          4&  25&   401K&   Class.\\
   \movie&          3&  29&   1M&   Class.\\
   \loyal&          4&  58&   118K&   Class.\\
   \ped&            3&  50&   18K&   Class.\\
   \event&          5&  129&  46K&  Class.\\
   \eventnot&       5&  129&  46K&  Class.\\
   \hline
   \fone&           9&  62&   133K&   Reg.\\
   \imdb&           7&  45&   203K&   Reg.\\
   \restbase&       3&  9&    28K&    Reg.\\
   \bio&            5&  11&   21K&   Reg.\\
    \midrule
    \bottomrule
    
  \end{tabular}
\end{table}

\noindent \noindent \textbf{\textit{Evaluation Metrics.}} Model performance evaluation is conducted using the following methodology: For classification tasks, primary metrics include \roc and \accuracy, complemented by \fscore and \average for a detailed performance analysis in parametric sensitivity studies. In regression tasks, evaluation relies on Mean Squared Error (\mae) and Mean Absolute Error (\mse).

In this research, we applied random sampling to the initial datasets, namely \olist, \event (\eventnot), \ped, and \loyal, because of their large size.




\noindent \textbf{\textit{Implementation.}}
In our experiment, we chose Python as the programming language. For the GNN part, we used the PyTorch-Geometric library, specifically adopting the GraphSAGE\cite{hamilton2017inductive} and GAT\cite{velivckovic2017gat} models. To encode the relational tables, we utilized the PyTorch Frame \cite{hu2024pytorch} library. Furthermore, training of the \xgboost and \lightgbm baseline models also directly used the interfaces provided by the PyTorch Frame library. Unless otherwise specified, the default threshold $\filterp$ for extracting important attribute relationships is set to 0.8. Each experimental trial was set with an 18-hour time limit, at which point the process would automatically conclude without achieving the results.

\begin{table*}
\small
  \caption{Results of the Classification Task (bold: the best; underline: the second best; Acc is \accuracy; AUC is \roc)}
  \label{tab:result_class}
  \begin{tabular}{c||cc|cc|cc|cc|cc|cc}
    \toprule
    \multirow{2}{*}{Method}   & \multicolumn{2}{c|}{\olist} & \multicolumn{2}{c|}{\movie} & \multicolumn{2}{c|}{\loyal} & \multicolumn{2}{c|}{\ped} & \multicolumn{2}{c|}{\event} & \multicolumn{2}{c}{\eventnot} \\
\cline{2-13}
       & Acc & \roc & Acc & \roc & Acc & \roc & Acc & \roc & Acc & \roc & Acc & \roc \\
\hline
    \midrule
    \texttt{\base}&                 0.852 & 0.8296&         0.7036& 0.5213&        0.5424& 0.5111&   0.515&  0.5257&    0.7323& 0.5022&            0.948&  0.5443\\
    \texttt{\bigtable}&             0.8339& 0.5616&         0.7517& 0.5001&        0.5353& 0.5146&   0.6069& 0.5842&    0.7456& 0.5701&     \textbf{0.9757}& 0.5344 \\
    \texttt{\randomselection}&      0.8335& 0.5113&         0.7216& 0.5547&        0.5203& 0.5034&   0.5344& 0.5437&    0.7396& \underline{0.6526}&  \underline{0.9736}& 0.5500 \\
    \texttt{\mutualinfor}&          0.8335& 0.5506&         0.7216& 0.5281&        0.5203& 0.5088&   0.6069& 0.5326&    \underline{0.7505}& 0.5001&            0.9729& 0.5165 \\
    \texttt{\backward}&             0.8527& 0.5001&         0.7525& 0.5001&        0.5421& 0.5001&   0.5336& 0.5007& \noresult & \noresult&        \noresult & \noresult \\ 
    \texttt{\xgboost}&              0.9068& 0.8852&         0.7011& 0.5988&        0.5421& 0.5575&  \underline{0.6625}& 0.5337&    0.7352& 0.6305&            0.9729& \underline{0.7691} \\
    \texttt{\lightgbm}&  \textbf{0.9130}& \textbf{0.8984}&  0.7051&  \underline{0.6245}&  0.5421& 0.5339&   0.6218& \underline{0.5834}&    0.7352& 0.6068&     0.9729& 0.6966 \\
    \texttt{\randomforest}&         0.8516& 0.8677&         0.7012& 0.6046&        0.5355& 0.5723&   0.6039& 0.5041&    0.7505& 0.6221&            0.9727& 0.5729 \\
    \texttt{\arda}&                 0.8526& 0.5003&         0.7011& 0.5048&        0.5421& 0.5078&   0.5276& 0.5002&    0.7395& 0.5023&            0.9729& 0.5020 \\
    \texttt{\ardano}&       0.8526& 0.5736& \underline{0.7543}& 0.5066&    \underline{0.5649}& \underline{0.5855}&   0.5461& 0.5007&    0.7352& 0.5173&            0.9332& 0.5049 \\
    \texttt{\leva}&                 0.8595& 0.5186&         0.7152& 0.5319&        0.52&   0.5105&   0.5525& 0.5528&    0.7396& 0.5&               0.9652& 0.5015 \\
    \texttt{\sys}&  \underline{0.9084}& \underline{0.8855}&  \textbf{0.7980}& \textbf{0.8432}&  \textbf{0.5825}& \textbf{0.6397}&   \textbf{0.685}& \textbf{0.7462}& \textbf{0.7516}& \textbf{0.6918}&     0.9603& \textbf{0.9233} \\
    \bottomrule 
  \end{tabular}
\end{table*}

\begin{table*}
\small
  \caption{Results of the Regress Task (bold: the best; underline: the second best)}
  \label{tab:result_reg}
  \begin{tabular}{c||cc|cc|cc|cc}
    \toprule
    \multirow{2}{*}{Method} & \multicolumn{2}{c|}{\fone} & \multicolumn{2}{c|}{\imdb} & \multicolumn{2}{c|}{\restbase} & \multicolumn{2}{c}{\bio}  \\
\cline{2-9}
       & \mae & \mse & \mae & \mse& \mae & \mse & \mae & \mse \\
\hline
    \midrule
    \texttt{\base}&                  489.878& 248807.529 &     6.935&       50.563&         0.445&       0.341&        \underline{0.925}&   1.316      \\
    \texttt{\bigtable}&              102.814& 68337.160  &     1.539&        4.578&         0.442&       0.345&        1.048&   1.732      \\
    \texttt{\randomselection}&       306.206& 113842.427 &     1.546&        3.975&         0.406&       0.323&        1.462&   3.449      \\
    \texttt{\mutualinfor}&           6.734&   66.279     &     1.224&        2.406&         0.401&       0.278&        1.80&    4.144  \\
    \texttt{\backward}&              15.048&  350.338    &     1.247&        2.453&         0.411&       0.265&        3.409&   12.875 \\
    \texttt{\xgboost}&               3.129&   13.913     & \noresult&    \noresult&         0.342&       0.306&        1.241&   2.205  \\
    \texttt{\lightgbm}&              3.109&   \underline{13.679}     & \underline{1.199}& 2.317&         0.336&       0.320&        1.114&   \underline{1.177}     \\
    \texttt{\randomforest}&          \underline{2.788}&   14.334     & \noresult&    \noresult&         0.416&       \underline{0.239}&        1.213&   2.182  \\
    \texttt{\arda}&                  23.548&  1347.496   &     2.123&        5.623&         0.407&       0.269&        3.4&     13.457  \\   
    \texttt{\ardano}&        9.188&   103.613    &     1.246& \textbf{1.552}&       0.409&       0.268&        5.241&   31.73     \\
    \texttt{\leva}&                  4.864&   40.4845    &     1.339&        3.017&         \underline{0.302}&       0.247&        1.28&    3.727     \\
    \texttt{\sys}&          \textbf{ 0.912}&   \textbf{3.227}& \textbf{1.113}& \underline{2.311}& \textbf{0.275}& \textbf{0.230}& \textbf{0.521}&   \textbf{0.5826} \\
    \bottomrule
  \end{tabular}
\end{table*}

\subsection{Comparison with Baselines}

Here, we evaluate the performance of the \sys model against other baselines, focusing on classification and regression tasks. 

\noindent \textbf{Classification. } As illustrated in Table \ref{tab:result_class}, we assess a variety of baseline methods across multiple datasets and utilizing different models. We use \accuracy and \roc metrics to evaluate each dataset across different models. Accuracy provides a straightforward and easily understandable performance metric, while \roc complements it by being effective in situations with imbalanced data. 
Table \ref{tab:result_class} shows the effectiveness of different methods on the six datasets. the performance of the base methods, including Base Table, Big Table, and Random, was inferior to that of embedding-based methods (including \xgboost,  \lightgbm, and \randomforest) as well as the \sys method. The experimental results indicate that using only the Base Table generally yielded poor results, and in the \olist dataset, the \bigtable method performed even worse than the Base Table, with the \roc significantly dropping from 0.8296 to 0.5616. This is because, although the \bigtable method added more features through join operations, the unfiltered features introduced noise. Meanwhile, the all-table join operation might also lead to data drift issues. The performance of the Random method was unstable: in the \loyal dataset, its metrics were lower than those of the \base and \bigtable, while in the \eventnot dataset, it performed better than the \base. This indicates that randomly selecting features might yield useful features or might introduce noise, thus systematic feature selection is necessary.

Embedding-based methods (\xgboost, \lightgbm, and \randomforest) performed the best among all baselines. For example, in the \olist dataset, \lightgbm achieved the best results in both \accuracy and \roc metrics. The advantage of these methods lies in their ability to dynamically evaluate feature importance during training and gradually optimize the feature subset. However, the downside is the high computational cost: \xgboost (100 epochs) took 35,899.234 seconds, \lightgbm (100 epochs) took 225,340.247 seconds, and \sys (200 epochs) took 9,284.666 seconds.

The mutual information method based on filtering performed worse than the embedding methods in all experiments. For instance, in the Loyal dataset, the \roc of the mutual information method was only 0.5088, significantly lower than that of \xgboost (0.5575), \lightgbm (0.5339), and \randomforest (0.5723). This performance gap stems from two main factors: firstly, filtering methods evaluate features solely based on univariate statistics, completely ignoring the interactions among features; secondly, their feature selection process is independent of model training and does not obtain real-time feedback on prediction performance. It is particularly noteworthy that traditional filtering methods lack effective mechanisms to handle text-type features (which need to be encoded into $300$-dimensional vectors), further limiting their performance.

The \backward method based on wrappers, although iteratively referencing model feedback to select features, neither outperformed the mutual information method nor reached the level of embedding methods. This method has two main deficiencies: first, constrained by a greedy search strategy, it has difficulty fully exploring the feature space and is prone to local optima; second, similar to filtering methods, it faces technical bottlenecks when handling high-dimensional text features. These limitations prevent the \backward method from discovering the optimal feature combination, ultimately affecting overall prediction performance.

The performance of \arda and \ardano was inferior to embedding-based methods and our method in all datasets. This is attributable to three main factors: first, the original \ardano algorithm does not support text feature processing, and the modified \ardano version in the baseline experiments sacrificed some performance for this; second, \ardano avoided processing issues by removing text features but led to the loss of effective feature information; finally, although the Leva method achieved better \accuracy values than other baselines in most datasets due to its graph structure design, its performance improvement was limited by the inability to fully explore potential relationships among features.

Our proposed \sys solution outperformed all baseline methods in terms of \accuracy and \roc metrics in the Movie, Loyal, and PED datasets, and also demonstrated competitive advantages in other datasets. The technical advantages of \sys lie in two aspects: 1) deeply exploring the potential relationships among features within the same table and finely splitting auxiliary tables; 2) accurately modeling join relationships among tables through a weighted heterogeneous graph structure, thus achieving more precise feature enrichment. These mechanisms collectively ensure the outstanding performance of the \sys method.

\noindent \textbf{Regression. } Using the same baselines as the classification experiments to construct regression tasks, Table 3 presents the experimental results, with performance evaluation using \mae and \mse. The trend of results is consistent with the classification experiments. The basic methods, including \base, \bigtable, and \randomselection, were inferior to the embedding-based methods (including \xgboost, \lightgbm, and \randomforest) as well as the \sys method. Moreover, embedding-based methods outperformed filter-based (mutual information) and wrapper-based (backward elimination) methods across all datasets. \leva and \ardano were superior to most baseline methods but inferior to our method.

\noindent \textbf{Summary. }To achieve better predictive performance, the key factors are: 1) adding features to the base table; and 2) ensuring that the added features are useful and do not introduce noise. By exploring the potential relationships among features within the same table and using weighted heterogeneous graphs to capture table structure, the \sys method ensures superior performance compared to other baselines.

\subsection{Sensitivity of \sys}

\begin{figure}
    \centering
    \includegraphics[width=0.4\textwidth]{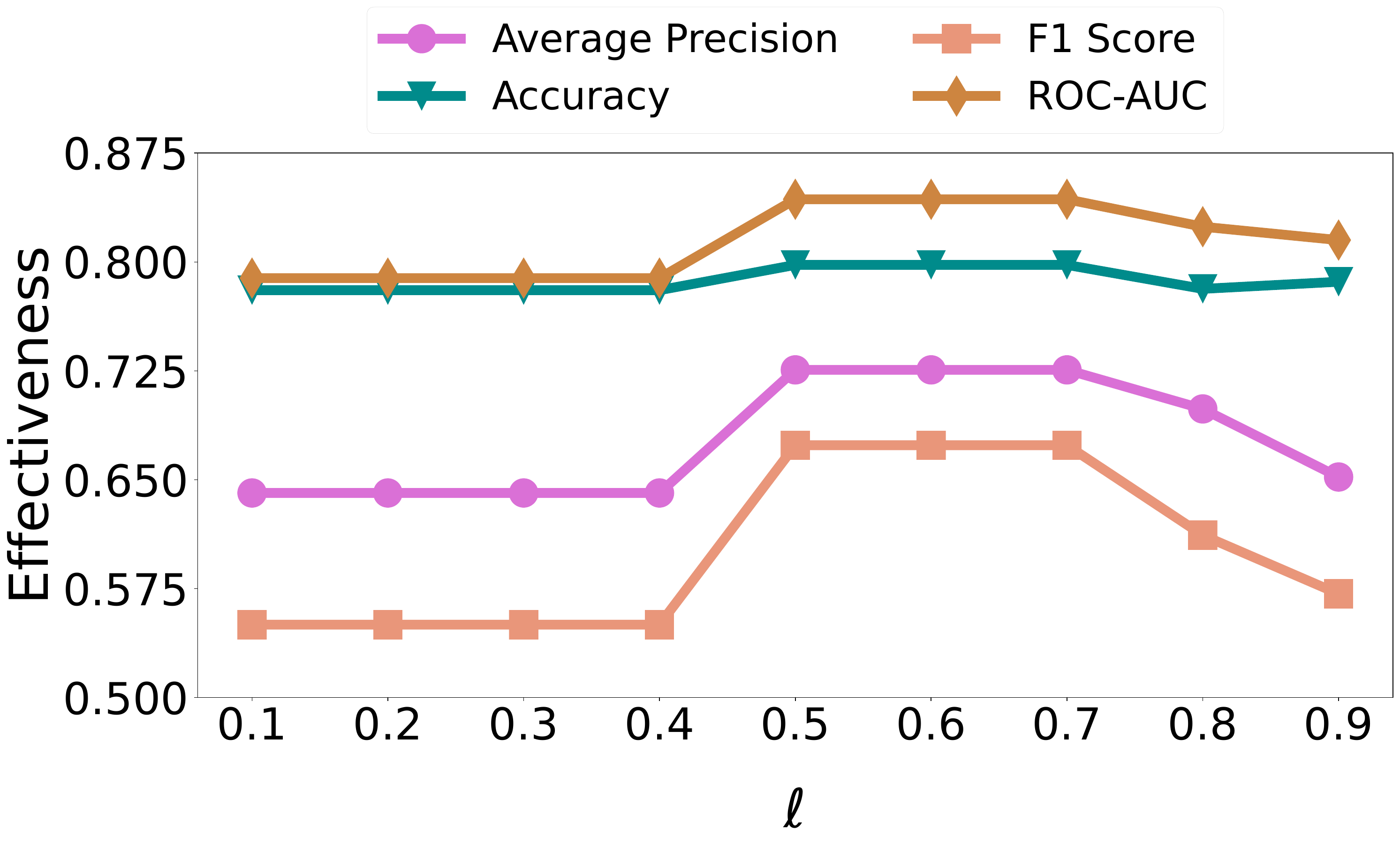}
    \caption{Performance of Different $\filterp$ on \movie Dataset}
    \label{fig:filter_param}
\end{figure}

We explore the impact of varying filtering parameters $\filterp$ on model performance within the \movie dataset, alongside the effect of $\filterp$ variations. As illustrated in Fig. \ref{fig:filter_param}, different $\filterp$ values influence model performance, depicted through a range of matrices. It is evident across these matrices that an increase in $\filterp$ initially strengthens and subsequently weakens the model's predictive capability. This occurs because, at a low $\filterp$, the relational dynamics between features are inadequately captured; as $\filterp$ increases, it begins to adeptly capture feature relationships within the same table. However, if $\filterp$ grows too large, it may excessively capture data, thereby reducing model efficacy. Notably, when $\filterp$ ranges from 0.1 to 0.4 and 0.5 to 0.7, all metrics are consistent due to the similar relationships captured by these parameter values.

As shown in Table \ref{tab:diff_graph_dis}, we can see that different graph discovery algorithms affect the performance of \sys on the \movie dataset. We compared two graph discovery algorithms, one is Complete Graph, and the other is Girvan-Newman(GN). The Girvan-Newman algorithm achieves community detection by continuously removing edges with high centrality, causing the network to gradually split into multiple tightly interconnected communities. At $\filterp =$  0.5 and 0.8, the Complete Graph algorithm performs better than the Girvan-Newman  algorithm, mainly because the Complete Graph algorithm is stricter in cluster discovery, discovering strong relationships between features and thereby reducing noise.

\begin{table}
\small
  \caption{Performance of Different Graph Discovery Algorithms on \movie Datasets (bold: the best)}
  \label{tab:diff_graph_dis}
  \begin{tabular}{c|cc|cc}
  \toprule
    \multirow{2}{*}{Method} & \multicolumn{2}{c|}{$\filterp=0.5$} & 
    \multicolumn{2}{c}{$\filterp=0.8$} \\
    \cline{2-5}
    & Complete & GN & Complete & GN \\
\hline
   \average&    \textbf{0.7257}&  0.7226&  \textbf{0.6988}&     0.6262   \\
   \accuracy&   \textbf{0.7980}&  0.7781&  \textbf{0.7815}&     0.7566   \\
   \fscore&     \textbf{0.6738}&  0.665&   \textbf{0.6118}&     0.5067   \\
   \roc&        \textbf{0.8432}&  0.8274&  \textbf{0.8243}&     0.7689   \\
     \bottomrule
  \end{tabular}
\end{table}

\subsection{Ablation Experiments}
\subsubsection{Attribute Relationship Mining.} 
\begin{figure}
    \centering
    \includegraphics[width=0.4\textwidth]{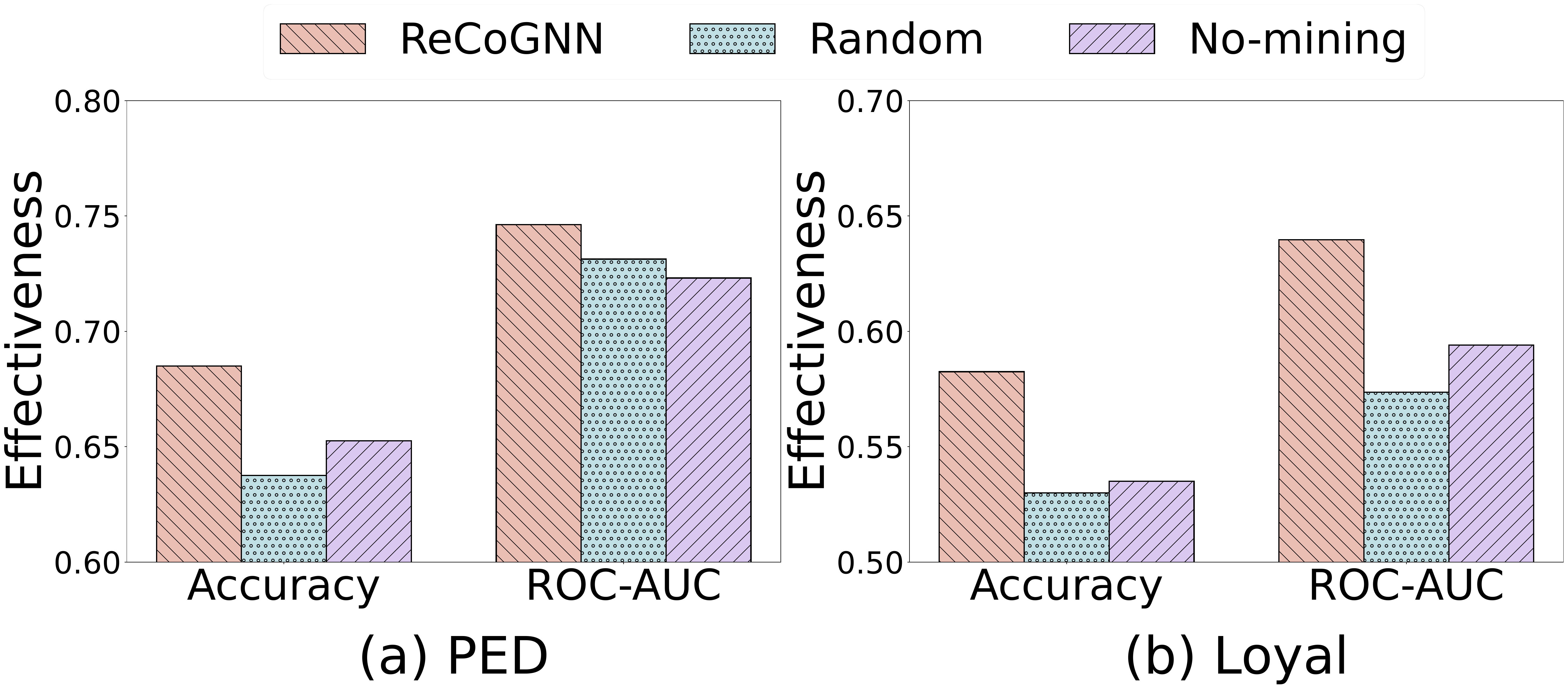}
    \caption{Ablation Studies of Attributes Relationship Discovery}
    \label{fig:abolish-split}
\end{figure}

Attribute Relationship Mining:
In the preceding discussion, we emphasized the importance of mining potential relationships between attributes in relational table data. Attributes with semantic or co-dependency relationships should be grouped together to preserve such information. Splitting tuples into node groups based on these relationships allows the graph to encode group-specific patterns. However, overly aggressive splitting may fragment coherent attributes, while insufficient mining risks merging unrelated attributes, both of which could degrade the performance of downstream tasks.
We designed an ablation study to evaluate the impact of attribute relationship mining. Three configurations were compared:
\begin{itemize}  [leftmargin=0.5cm]
    \item Graph-based mining: The method mentioned in Chapter 4, which detects correlations between attributes by statistically analyzing edge weight coefficients k.
    \item Random grouping: Attributes are split into random node pairs without relationship analysis.
    \item No mining: Each attribute is treated as an independent node.
\end{itemize}
Using \accuracy and \roc as evaluation metrics, we conducted experiments on the \ped and \loyal classification datasets. As shown in Figure \ref{fig:abolish-split}, the accuracy of the feature latent relationship mining model on the relationship table is significantly better than that of randomly combined feature relationships and without feature relationship mining.

\subsubsection{Edge Weights.} 
In section \ref{sec:feature_augmentation}, we discussed how to use weight graphs to enrich and predict the features of each tuple in the base table. To examine the impact of edge weights on predictive performance, we conduct an ablation study by modifying the graph's edge properties. In this study, we compare the results of models with and without edge weights to assess their influence on the overall accuracy. 
As shown in Fig. \ref{fig:abolish-edge}, in the \ped and \loyal datasets, the prediction accuracy on the weighted graphs surpasses that of the methods without introduced weights; the same significant effect is observed in regression data sets. This is because GNN can learn edge weights, effectively mitigating the impact of noisy features on the final prediction during the message-passing process, thereby enhancing the overall performance of the model.

\subsubsection{Graph Construction Schema.} 
\begin{figure}
    \centering
    \includegraphics[width=0.4\textwidth]{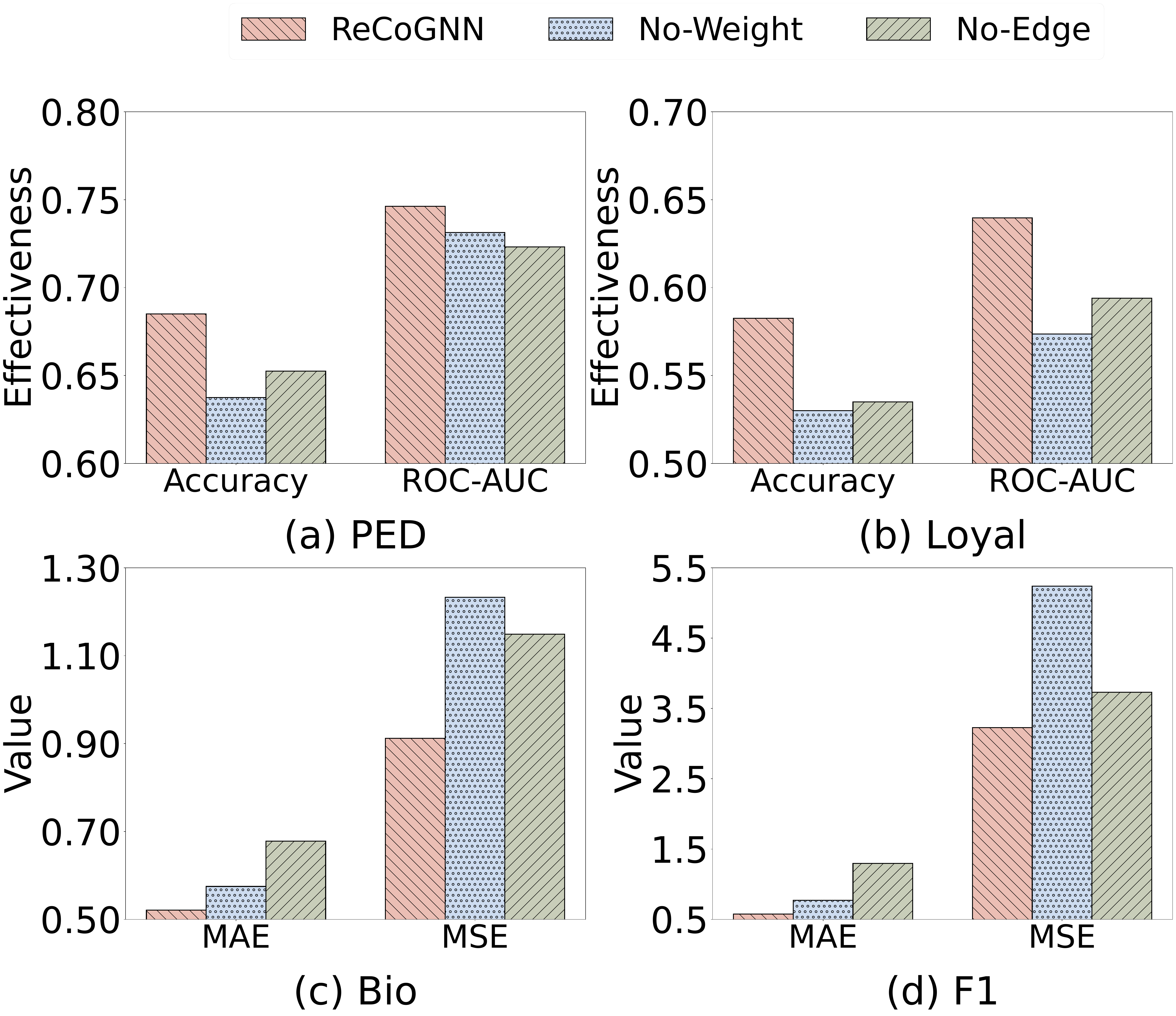}
    \caption{Ablation Studies on Edge Weights and Added Similarity Edges}
    \label{fig:abolish-edge}
\end{figure}

To enhance the prediction accuracy by using relational data, we convert multi-table datasets into graphs. The primary table, known as the base table, contains nodes with labels, while the other tables provide auxiliary information to enrich the feature space of the base table. The experiment focuses on the impact of retaining or removing similarity-based edges on prediction performance. As depicted in Fig. \ref{fig:framework}, we establish two types of edges in our graph construction: (1) join relations between tables, and (2) similarity-based edges within the base table. For the similarity-based edges, each row in the base table is encoded into an embedding, and edges are added between tuples with high similarity scores. To evaluate the impact of these similarity edges, we conduct two sets of experiments: one with similarity edges included and another with them removed.
From the results shown in Fig \ref{fig:abolish-edge}, it is evident that retaining similarity-based edges in the graph significantly boosts the predictive accuracy or enhances the model's prediction precision. This is because introducing similarity edges can enhance the flow of information between nodes from the base table. More efficient information transmission can prompt the model to learn more optimized node embeddings. This greatly benefits downstream tasks such as classification and regression.

\section{Related Work}
\label{sec:Related}

\noindent \textbf{Feature Augmentation} Several notable works have addressed feature augmentation from relational tables. Early studies like \cite{kumar2016join, shah2017key} avoided unnecessary joins by leveraging foreign key constraints. Later, ARDA \cite{chepurko2020arda} proposed an end-to-end automated system with heuristic feature selection, while AutoFeature \cite{liu2022feature} introduced reinforcement learning for exploration-exploitation trade-offs. Autofeat \cite{ionescu2024autofeat} extended this by discovering transitive features through multi-hop joins, albeit with increased computational overhead. Further improvements include METAM's \cite{galhotra2023metam} goal-oriented discovery framework and FEATPILOT's \cite{liang2025featpilot} efficient multi-hop augmentation via clustering and LSTM-based prediction. Alternatively, \cite{wang2022coresets} adopted coreset-based selection to bypass table materialization while preserving accuracy.

\vspace{\vspacelen pt} 
\noindent \textbf{Modeling Attribute Relationships in Tabular Data}
Attribute interaction methods are classified into Factorization Machines (FM) \cite{rendle2010FM}, their enhancements, deep learning-based models, and recent GNN-based approaches. FM traditionally handle second-order feature interactions using vector inner products. Enhanced versions like Field-aware Factorization Machines \cite{juan2016FFM} and Attentional Factorization Machines \cite{xiao2017AFM} improve interaction modeling. Deep learning approaches, such as DeepFM \cite{guo2017deepfm} and xDeepFM \cite{lian2018xdeepfm}, merge FM and deep networks to capture explicit and implicit high-order interactions. GNNs excel at modeling complex feature relationships; Fi-GNN \cite{li2019fignn}, for instance, treats features as nodes and interactions as edges for versatile attribute representation.

\vspace{\vspacelen pt} 
\noindent \textbf{GNN for Tabular Data}
In recent research, Graph Neural Network (GNN) models have become prominent for processing relational data, expertly crafted to handle graph-structured information. These models iteratively refine node embeddings by integrating data from both the nodes and their neighbors, using these enhanced representations for predictions. Key contributions in linking relational data with GNNs are from Schlichtkrull et al. \cite{schlichtkrull2018modeling}, Cvitkovic et al. \cite{cvitkovic2020supervised}, and Zahradník et al. \cite{zahradnik2023deep}. GNN models can be categorized into inductive and transductive types based on their capability to manage unseen nodes. Inductive GNNs, such as GraphSAGE \cite{hamilton2017inductive}, can generalize to new nodes or graphs. In contrast, transductive GNNs like GCN \cite{kipf2016semi} and GIN \cite{xu2018gin} optimize node representations using the full graph but cannot adapt to new graphs. A hybrid model is GAT \cite{velivckovic2017gat}, which supports both approaches. Our model employs the inductive strategy, allowing it to accommodate newly introduced samples in future scenarios.

\section{Conclusion}
\label{sec:conclusion}

In this paper,we introduce \sys, an autonomous feature augmentation framework. The framework employs a two-stage progressive column relationships discovery mechanism:(1) \sys performs parallelized modeling of latent inter-column dependencies within each auxiliary table, followed by partitioning each auxiliary table into semantically coherent segments with enhanced attribute correlations. (2) In the next stage, we construct a weighted heterogeneous graph incorporating both the base table and segmented data, enabling the GNN's message passing to perform single-pass feature selection and base table feature augmentation. We conducted comprehensive experiments on real-world datasets to evaluate the performance of \sys against other baseline methods. The results demonstrate that \sys consistently outperforms the baselines and exhibits superior performance.


\clearpage

\balance
\bibliographystyle{ACM-Reference-Format}
\bibliography{DA.bib}

\end{document}